\documentclass[english]{article}
\usepackage[T1]{fontenc}
\usepackage[latin9]{inputenc}
\usepackage{amssymb}

\makeatletter
\newcommand{\lyxaddress}[1]{
\par {\raggedright #1
\vspace{1.4em}
\noindent\par}
}

\makeatother

\usepackage{babel}
\begin{document}
\def\lambdabar{\lambda\kern-1ex\raise0.65ex\hbox{-}}

\title{A Short Note on Minimal Length}

\author{Md. Moniruzzaman\thanks{Permanent Address: Department of Physics, Mawlana Bhashani Science
and Technology University, Santosh, Tangail-1902, Bangladesh}~ and S. B. Faruque\thanks{Corresponding Author: Email: awsbf62@yahoo.com}}

\maketitle

\lyxaddress{Department of Physics, Shahjalal University of Science and Technology,
Sylhet 3114, Bangladesh }
\begin{abstract}
After revival of the concept of minimal length, many investigations
have been devoted, in literature, to estimate upperbound on minimal
length for systems like hydrogen atom, deuteron etc. We report here
a possible origin of minimal length for atomic and nuclear systems
which is connected with the fundamental interaction strength and the
Compton wavelength. The formula we appear at is numerically close
to the upperbounds found in literature.
\end{abstract}

\section{Introduction }

The pursuit to know completely the nature of space-time in modern
times dates back to the time of Einstein\textquoteright s and Heisenberg\textquoteright s
epic works. Heisenberg first noticed that a non-zero minimal length
is inevitable, but he could not incorporate it in a consistent theory
{[}1{]}. But the history of a minimal length is becoming rich day
by day. In the last decade of the 20th century, it became clear how
a minimal length appears in quantum gravity {[}2{]}, string theory
{[}3{]}, etc., and through the works of Kempf et al {[}4-6{]}, a deformed
quantum mechanics appeared which incorporates a non-zero minimal length
within a generalized uncertainty principle (GUP). The generalized
uncertainty principle (GUP) is usually expressed as
\begin{equation}
\triangle x\triangle p\geqslant\frac{\hbar}{2}\left[1+\beta\left(\triangle p\right)^{2}\right]
\end{equation}
where $\beta$ is a small parameter. Equation (1) leads to the minimal
length $\triangle x_{min}=\hbar\sqrt{\beta}$. Originally, this minimal
length was envisaged as a gravitational induced uncertainty and the
scale of this length was usually understood to be of the order of
Planck length$\left(10^{-35}m\right).$ But Eq.(1) contains the UV/IR
mixing, whereby it is thought that the minimal length can be much
larger than the Planck length. 

Brau {[}7{]} calculated the effect of a minimal length on the spectrum
of the hydrogen atom and estimated an upperbound on minimal length
inherent in H-atom which is of the order of $10^{-17}m$. Thereafter,
many authors devoted time in estimating upperbound on minimal length
inherent in many quantum systems and a large varity of results appeared.
However, for hydrogen-like systems, the estimated value of minimal
length falls around the value found by Brau. We have previously estimated
{[}8{]} the upperbound on minimal length using deuteron quadrupole
moment, which is about $10^{-16}m$.

Due to this development of the matter, some authors indicated that
minimal length may depend on the systems under consideration . Moreover,
Kempf {[}9{]} revealed that non-pointness of a particle considerably
affects energy spectrum of systems like harmonic oscillator, and Sastry
{[}10{]} presented a theory of extended quantum particles where non-pointness
of particles is shown to lead to non-zero minimal uncertainty in position.
Influenced by these works, we are going to show here that a minimal
length could be associated with every quantum system because of the
non-pointness of the particles manifested by their being part of a
system and by how their positions are measured. These two ingredients
are inseparably associated with every quantum system and with the
conceptual framework of Heisenberg uncertainty principle. And one
obviously finds a minimal length to be operational in every quantum
system when non-pointness of particles whose positions are measured
is considered within the framework of quantum measurement. Through
due consideration of non-pointness of particles we arrive at a formula
for minimal length. The work we present here is original and of fundamental
importance for quantum physics. The paper is organized as follows:
In Section 2, we derive the formula for minimal length and present
some numerical values. In Section 3, we conclude.

\section{Minimal Uncertainty in Position }

According to Heisenberg\textquoteright s original thought experiment
{[}11{]}, a particle\textquoteright s position is measured through
the interaction of a photon with the particle. The photon gets scattered
and the scattered photon is viewed through a microscope which immediately
leads to the ordinary uncertainty relation 
\begin{equation}
\triangle x\triangle p\geqslant\frac{\hbar}{2}
\end{equation}
Here, however, the particle, say electron, is considered as a point
particle. But the interaction which is used to detect the electron
always finds the particle as non- pointlike. This is manifest through
the finite cross-section of the interaction which is always like Compton
scattering. Every impingement reveals a finite size of the electron
given by, the low-energy limit of Compton scattering cross-section{[}12{]}
\begin{equation}
\sigma=\frac{8\pi}{3}\left(\frac{e^{2}}{\hbar c}\frac{\hbar}{m_{e}c}\right)^{2}
\end{equation}
where $e$ is the charge of the electron and $m_{e}$ is the mass
of electron. This is actually the Thomson scattering cross-section,
the low energy limit of the Compton cross-section. So, a measure of
non-pointness of the electron associated with Eq.(3) is given by a
minimal length,

\begin{equation}
\triangle x_{min}\thickapprox\frac{e^{2}}{\hbar c}\frac{\hbar}{m_{e}c}
\end{equation}

\[
=\alpha\lambdabar_{c}
\]
where $\alpha$ is the fine structure constant and $\lambdabar_{c}$
is the reduced Compton wavelength of the electron.

Next, consider the H-atom. When energy of H-atom is considered, the
particle whose energy is measured is electron and its size is linked
with the size of the orbit. In the ground state , the size of the
electron, $\frac{\hbar}{m_{e}c}=\lambdabar_{c}$ and the radius of
the orbit, Bohr radius $a_{0}$ are linked by
\begin{equation}
\frac{\lambdabar_{c}}{a_{0}}=\alpha\backsimeq\frac{1}{137}
\end{equation}
If the electron is in a higher excited state the ratio $\frac{\lambdabar_{c}}{r_{n}}$
gradually approaches zero. Here lies another measure of non-pointness
of the electron. This measure here is largest in the ground state.
So, minimal size of the electron or it\textquoteright s non-pointness
is not fixed if we consider the system where it is. The measure of
this non-pointness varies between the minimum zero, which says it
is really pointlike, and the maximum, $\alpha$, which says it is
somewhat non-pointlike. But as long as we consider the electron in
the H-atom this measure is never zero. For other particles in other
systems, the measure of this non-pointness may turn out bigger. For
example, in deuteron, the measure of non-pointness is more and can
be taken to be $\alpha_{s}\thickapprox0.2$, where $\alpha_{s}$ is
the coupling constant of strong interaction. We may find $\alpha_{s}$
by dividing the Compton wavelength of the reduced mass of neutron-proton
system by the size of deuteron. The result is close to the value$\left(\alpha_{s}\thickapprox0.1\right)$
in quantum field theory {[}12{]}.

Now, in literature, the systems for which the minimal length are being
reported are bound systems like harmonic oscillator, hydrogen atom,
deuteron, etc. Orginally, the minimal length appeared for systems
that are in Planck scale. For all such systems, the minimal length
can be linked with the two pieces of non-pointness discussed above.
We can fairly assume the minimal length to be proportional to both
of those pieces. That is, $\triangle x_{min}\propto\alpha$ and $\triangle x_{min}\propto\left(\alpha\lambdabar_{c}\right)$,
where $\alpha$ now represent the coupling strength of the interaction
by which the particle under consideration is bound within the system,
and $\lambdabar_{c}$ is the reduced Compton wavelength of the particle.
We thus write 
\begin{equation}
\triangle x_{min}\thickapprox\alpha^{2}\lambdabar_{c}
\end{equation}

For hydrogen, $\alpha\backsimeq\frac{1}{137}$ and $\lambdabar_{c}=\frac{\hbar}{m_{e}c}=3.86\times10^{-13}m$
and $\triangle x_{min}$ turns to be about $2\times10^{-17}m$, which
is close to Brau\textquoteright s value {[}7{]}. For deuteron, $\alpha_{s}\thickapprox0.2$
and $\lambdabar_{c}\backsimeq\frac{2\hbar}{m_{p}c}=4.19\times10^{-16}m$
, where $\frac{m_{p}}{2}=8.35\times10^{-28}kg$ is the reduced mass
of deuteron. Thus, for deuteron, $\triangle x_{min}$ turns out to
be about $1.67\times10^{-17}m$, which is close to our value {[}8{]}.
For Planck scale system, $\lambdabar_{c}\approx10^{-35}m$ and $\alpha_{p}\approx1$
and the minimal length turns out, as expected, to be the Planck length. 

Thus, we arrive at a value for minimal length for bound systems as
given by Eq.(6) and get some confirmation of the formula (6) from
literature as quoted above. Strictly speaking, formula (6) represent
the upperbound on minimal length for bound systems and can be linked
with the uncertainties of dynamical properties of such systems. However,
a minimal length for every quantum system is inevitable which is clear
from the discussion above and this minimal length can be fairy considered
to be not more than that given by Eq.(6). For systems bound by fundamental
interaction, Eq.(6) is easy to evaluate, but for other systems, further
investigation is necessary.

\section{Conclusion}

Minimal length, first appeared in string theory, quantum gravity,
and related fields, is the minimal uncertainty in position measurement
of a quantum particle. In this paper, we have argued that a minimal
length is inevitable in Heisenberg\textquoteright s uncertainty principle.
The long silence about minimal length after Heisenberg\textquoteright s
original formulation of the uncertainty principle, is due to its very
small size. The most fundamental minimal length is of the Planck scale,
but in bound systems like hydrogen atom, deuteron etc., the minimal
length can appear to be much bigger. A minimal length can be considered
to be operational in every quantum system, which is shown in this
work to be linked with non-pointness of particles like electron, nucleon,
etc.. The non-pointness discussed in this paper has two pieces. One
is linked with the particles being part of a bound system and the
other is linked with the method of measurement of the particle\textquoteright s
position. The measure of non-pointness, when we consider a particle
within a quantum system, is given approximately by the coupling strength
of the interaction by which the particle is bound within the system.
The other measure of non-pointness, when we consider the way of position
measurement, is given approximately by the square root of interaction
cross section, the interaction being like Compton scattering. We thus
arrived at the formula (6) for minimal length whose numerical value
is close to the upperbounds on minimal length found in literature.
Therefore, we find a significant piece of physics in this work. Further
investigation on this particular orgin of minimal length in quantum
systems may shed more light on the topic.

\end{document}